\documentclass[journal]{IEEEtran}

\ifCLASSINFOpdf
\else
   \usepackage[dvips]{graphicx}
\fi
\usepackage{url}

\usepackage{graphicx}
\usepackage{amsmath,amssymb,amsfonts}
\usepackage[compact]{titlesec}  %%spaces between paragraphs, sections..
\titlespacing{\section}{0pt}{*0}{*0}

\begin{document}

\title{CNN based Channel Estimation using NOMA for mmWave Massive MIMO System }

\author{T S Anu, Tara Raveendran
%\author{First A. Author, \IEEEmembership{Fellow, IEEE}, Second B. Author, and Third C. Author, Jr., \IEEEmembership{Member, IEEE}
\thanks{This paragraph of the first footnote will contain the date on which you submitted your paper for review. }
\thanks{T S Anu is with the Department of Electronics and Communication Engineering, College of Engineering Trivandrum, Thiruvananathapuram 695016, India (e-mail: tsanu@cet.ac.in).}
\thanks{Tara Raveendran is with the Department of Applied Electronics and Instrumentation Engineering, Government Engineering College Kozhikode 673005 (e-mail: tara.rajesh1@gmail.com).}}

%\markboth{Journal of \LaTeX\ Class Files, Vol. 14, No. 8, August 2015}
\maketitle

\begin{abstract}
Non-Orthogonal Multiple Access (NOMA) schemes are being actively explored to address some of the major challenges in 5th Generation (5G) Wireless communications. Channel estimation is exceptionally challenging in scenarios where NOMA schemes are integrated with millimeter wave (mmWave) massive multiple-input multiple-output (MIMO) systems.  An accurate estimation of the channel is essential in exploiting the benefits of the pairing of the duo-NOMA and mmWave. This paper proposes a convolutional neural network (CNN) based approach to estimate the channel for NOMA based millimeter wave (mmWave) massive multiple-input multiple-output (MIMO) systems built on a hybrid architecture. Initially, users are grouped into different clusters based on their channel gains and beamforming technique is performed to maximize the signal in the direction of desired cluster. A coarse estimation of the channel is first made from the received signal and this estimate is given as the input to CNN to fine estimate the channel coefficients. Numerical illustrations show that the proposed method outperforms least square (LS) estimate, minimum mean square error (MMSE) estimate and are close to the Cramer-Rao Bound (CRB).
\end{abstract}


\IEEEpeerreviewmaketitle

\section{Introduction}
Recently, millimeter wave (mmWave) communication systems are emerging as a promising technology for next generation wireless communications [1]. Current wireless systems use sub-6 GHz spectrum (300MHz–6GHz). Only limited bands are available in sub-6 GHz spectrum. mmWave exploits the high frequency mmWave band (6–300GHz) where the spectrum is less crowded and under-utilized. \\
%Large coverage area and mobility in non-line-of-sight (NLOS) environments are the two barriers that keep off to use this under-utilized band. Beamforming with large antenna arrays can compensate high propagation loss in mmWave spectrum by extending the coverage to a longer range [5]. Hybrid analog–digital architecture is used to reduce the number of radio frequency (RF) chains as it is difficult to equip RF chain on each antenna due to area of chip, complexity and cost [7]. \\
Non-Orthogonal multiple access (NOMA) is one of the key principles for radio access in 5G and future networks [2]. NOMA is based on the idea that multiple users are multiplexed over the same orthogonal resource block (time slot, subcarrier etc.). Eventhough there is a huge amount of spectrum available in mmWave band, the use of NOMA in mmWave band is important due to: (i) NOMA with mmWave network can enable mass connectivity. (ii) The spectral efficiency of NOMA is higher compared to Orthogonal Multiple Access (OMA). (iii) NOMA can also lead to very low latency via grant free transmission.\\
Accurate channel state information (CSI) is important to get full benefits of mmWave with NOMA technology. However, to obtain accurate CSI is considered as one of the main challenges. A new linear estimator is introduced in [3] to estimate the channel by maximizing the signal to interference and noise (SINR) ratio of strong user and ensuring limited SINR for the weak user. In [4], singular value decomposition (SVD)  with filter precoding is used to estimate the channel. For each channel path, the iterative algorithm in [5] acquire channel gain and direction of arrival to find channel estimate.\\
Deep learning (DL) has been gaining increasing popularity in recent years. Deep learning is a data driven method where the network are developed over a large number of training dataset and well defined mathematical model is nonessential. This features leads to utilize deep learning in communication system [6]. A learned denoising  based approximate message passing (LDAMP) network is used to estimate the channel for mmWave massive MIMO system in [7]. A DL based algorithm called ChannelNet is proposed in [8] where the frequency response of the fading channel is considered as an image and applied to super resolution network followed by image restoration network to estimate the channel. A deep neural network (DNN) followed by bidirectional long short-term memory (Bi-LSTM) is proposed in [9] to estimate and track the mmWave channel. \\
A conventional channel estimation method has a tractable mathematical model. However, practical channel may have imperfections and more complicated and the model may perform poorly. On the other hand, CNN can learn features associated with the underlying channel from large number of training data and provides good estimate of channel which is more accurate and less complex. Firstly users undergo clustering based on channel gains and each NOMA cluster will have two users. Beamforming is performed to cancel out the interference from other clusters. A CNN based approach is proposed to estimate the channel of each users in NOMA clusters. Numerical Illustration and results shows that the proposed method outperforms the least square (LS) estimator, minimum mean square error (MMSE) estimator and close to Cramer-Rao Bound (CRB).\\
The paper is organized as follows. Section 2 gives the system model and brief introduction to NOMA model followed by proposed CNN based channel estimation in Section 3. Numerical Illustration and Results are provided in Section 4 and concluding remarks in Section 5.\\

\section{System Model}
In this section, channel model, basic NOMA model, clustering and beamforming techniques are discussed in detail.
\subsection{Channel Model} 
Consider a mmWave massive MIMO system, where the base station (BS) with $N_t$ antennas and $N_t^{RF}$ RF chains transmits signals to a single user with $N_r$
antennas and $N_r^{RF}$ RF chains. Phase shifters are employed to connect a large number of antennas with a much fewer number of RF chains at both the BS and the user sides. We therefore assume $N_t$ $\gg$  $N_t^{RF}$ and $N_r$ $\gg$  $N_r^{RF}$.\\
According to [10], the $N_r$ $\times$ $N_t$ channel matrix between the
receiver and the transmitter in the delay domain is given by:
\begin{equation}
\mathbf{H}(\tau) =  \sum_{l=1}^{L}\alpha_l \delta(\tau - \tau_l)a_R(\varphi_l)a_T^H(\phi_l)
\end{equation}
where $L$ is the number of main paths, $\alpha_l \sim C\mathcal{N}(0,\sigma^2_\alpha)$ is the propagation gain of the $l$th path with $\sigma^2_\alpha$ being the average power
gain, $\tau_l$ is the delay of the $l$th path, $\varphi$ and $\phi \in$ [0,2$\pi$] are the azimuth angles of arrival and departure (AoA/AoD) at the receiver and the transmitter, respectively. For uniform linear array (ULA), the corresponding response vectors can be expressed as:
\begin{equation}
a_R(\varphi) = \frac{1}{\sqrt{N_r}} [1,e^{-j2\pi\frac{d}{\lambda}\sin{\varphi_l}},...,e^{-j2\pi\frac{d}{\lambda}(N_r-1)\sin{\varphi_l}}]^T
\end{equation}
\begin{equation}
a_T(\phi) = \frac{1}{\sqrt{N_t}} [1,e^{-j2\pi\frac{d}{\lambda}\sin{\phi_l}},...,e^{-j2\pi\frac{d}{\lambda}(N_t-1)\sin{\phi_l}}]^T
\end{equation}
where $d$ and $\lambda$ denote the distance between the adjacent antennas
and carrier wavelength respectively. A brief introduction of the basic NOMA model is presented next.

\subsection{Basic NOMA Model}
In power domain (PD) NOMA, multiple users serve at same orthogonal resource block but allocated different power levels. Users with better channel conditions are allocated less power and vive-versa. Fig. 1. shows NOMA scheme via power domain multiplexing. \\
\begin{figure}[h]
	\centering
	\includegraphics[width=1\linewidth]{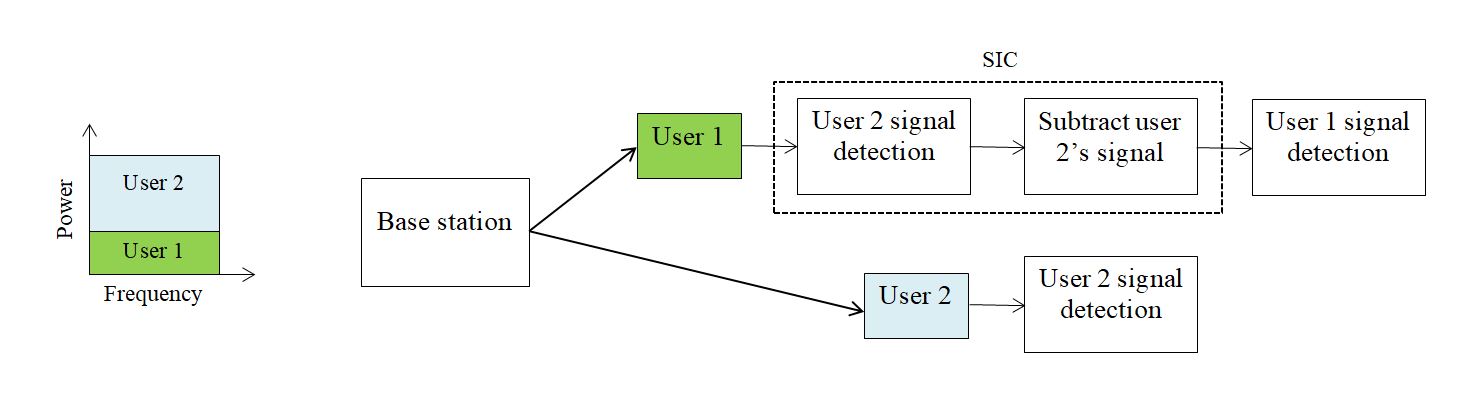} 
	\caption{Illustration of downlink NOMA transmission via power domain multiplexing}
	%\label{fig:lena}
\end{figure}
Consider two user (User 1 and User 2) scenario and assume User 1 has strong channel compared to User 2 i.e., $|h_1| > |h_2|$. So less power will be allocated to User 1 and more power to User 2. At the receiver, User 2 decodes signal first from the received signal treating User 1 signal as noise. However, for User 1 direct detection is not possible since more power is allocated to User 2 and hence large interference User 2. Therefore, User 1 detect and decode User 2 signal first and then use Successive Interference Cancellation (SIC) to cancel it. After SIC, User 1 detects its own signal from noise. For this process, channel state information (CSI) needs to be known at base station and user. Partitioning of users into different clusters is described next.

\subsection{Dynamic Clustering}

A low complexity user grouping scheme is proposed in [11]. It exploits channel gain difference among users in a NOMA cluster and group them into single or multiple clusters to enhance the sum throughput of the system. Clustering in downlink NOMA pairs highest channel gain user and lowest channel gain user into the same NOMA cluster while the second highest channel gain user and the second lowest channel gain user into another cluster and so on. High channel gain users are paired with low channel gain users to increase the throughput.
The reason is that the high channel gain users can achieve a higher rate with low power levels while making a large fraction of power available for weak channel users. $N_r$ users are grouped into $N_c$ cluster with each cluster having two users (one strong channel user and a weak channel user).

The channel associated with users in each cluster are grouped to form channel matrix $\mathbf{H_{c}} \in \mathbb{C}^{N_r \times N_t} $ given by:
\begin{equation}
\mathbf{H_{c} = [H_1 \hspace{.2cm}H_2 \hspace{.2cm} ... \hspace{.2cm}H_{N_c}]^T}
\end{equation}
where each $\mathbf{H_k = [h_i \hspace{.2cm} h_j]^T}; \mathbf{k= 1,2,...N_c}$ consisting of a strong and a weak channel user. Beamforming technique is used to cancel out interference from other clusters which is described next.

\subsection{Beamforming}
To mitigate the interference from other cluster (inter cluster interference), beamforming need to be done. Beamforming maximizes the signal in the direction of desired cluster and place a null in the direction of interference. Beams from one cluster are made orthogonal to the channel of users in other clusters and thereby suppressing interference [12]. \\
The clustered channel matrix $\mathbf{H_{c}}$ is given by (4). Let $\mathbf{h_i}$ and  $\mathbf{h_j}$ be the strong and weak channel user in cluster $\mathbf{k}$. Therefore, $\mathbf{H_k = [h_i \hspace{.2cm} h_j]^T}$.  Form a submatrix by removing channel coefficients of users in the cluster $\mathbf{k}$ as:
\begin{equation}
\mathbf{H_{-k} = [H_1 \hspace{0.2cm} H_2 \hspace{0.2cm} ...\hspace{0.2cm}H_{k-1} \hspace{0.2cm} H_{k+1} \hspace{0.2cm} . . . \hspace{0.2cm} H_{N_c}]^T} \in \mathbb{C}^{(N_r -2) \times N_t}
\end{equation}
The orthogonal projection or the beamforming of $\mathbf{H_{-k}}$ is given by:
\begin{equation}
\mathbf{B_k = I_{N_t} - H_{-k}^H (H_{-k}H_{-k}^H)^{-1}H_{-k}}
\end{equation}
This beamforming matrix $\mathbf{B_k}$ will be orthogonal to channels of users in other cluster i.e.,
\begin{equation}
\mathbf{B_k \hspace{0.2cm} h_l = 0 \; ;\; l \neq i,j}\\
\end{equation}
The next section presents the framework for CNN based channel estimation for mmWave MIMO system. \\

\section{CNN Based Channel Estimation}
In this section, algorithm and CNN architecture are discussed. 
\subsection{Algorithm Description}
Fig. 2. shows procedure for CNN based channel estimation.
\begin{figure}[h]
	\centering
	\includegraphics[width=1\linewidth]{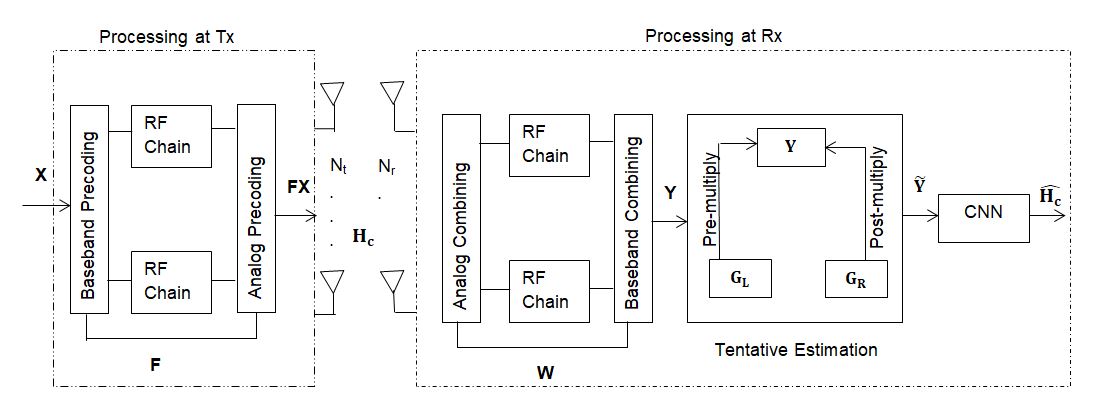} 
	\caption{CNN based channel estimation}
	%\label{fig:lena}
\end{figure}
The pilot signal matrix at the receiver for cluster $\mathbf{k}$ is given by:
\begin{equation}
\mathbf{Y_k}= \mathbf{W^H H_kB_kFS + \tilde N}
\end{equation}
where $\mathbf {W =[w_1,...w_{M_r}]}$ and $\mathbf {F =[f_1,...f_{M_t}]}$ are combining matrix and precoding matrix respectively. $\mathbf {M_r}$ and $\mathbf {M_t}$ are the numbers of combining and precoding vectors. $\mathbf {H_k}$ and $\mathbf {B_k}$ are the channel matrix and beamforming matrix of cluster $\mathbf{k}$. $\mathbf{\tilde N = W^H N}$ denotes the effective noise after combining and $\mathbf{N}$ is additive white Gaussian noise (AWGN) with
$C\mathcal{N}(0,1)$ elements before combining. $\mathbf{S}$ is the transmitted pilot symbols which is a $N_t \times N_t$ diagonal matrix given by:
\begin{equation}
\mathbf{S} = \sqrt{\alpha P + (1-\alpha)P} \; \mathbf{I}
\end{equation}
where $P$ is the total transmit power and $\alpha$ is the power scaling coefficient. The received signal goes through tentative estimation module [13] and a coarse estimate of $\mathbf {H_c}$ is obtained i.e.,
\begin{equation}
\mathbf{\tilde{Y}_k = G_L Y_k G_R}
\end{equation}
where, 
\begin{equation}
\mathbf{G_L} =
\begin{cases}
\mathbf{W}, & M_r < N_r\\
\mathbf{(WW^H)^{-1} W}, & M_r \geq N_r
\end{cases} 
\end{equation}
and
\begin{equation}
\mathbf{G_R} =
\begin{cases}
\mathbf{F^H}, & M_t < N_t\\
\mathbf{F^H(FF^H)^{-1}}, & M_t \geq N_t
\end{cases} 
\end{equation}
The tentatively estimated signal $\mathbf{\tilde{Y}_k}$ of each cluster are then grouped to form a $N_r \times N_t$ matrix given by:
\begin{equation}
\mathbf{\tilde Y = [\tilde{Y}_1,\hspace{0.2cm}...\hspace{0.2cm}\tilde{Y}_{N_c}]^T}
\end{equation}
The received signal matrix $\mathbf{\tilde Y}$ are then input to CNN, which outputs the estimated channel matrices $\mathbf{\hat H_c}$. The architecture of CNN and training process is described next.

\subsection{CNN Training}
The training data is generated according to channel model described in (1) and undergoes clustering to select users in a NOMA cluster with $(\mathbf{\tilde{Y}_i,H_{c_i}})$ denoting the $i$th sample, where $\mathbf{\tilde{Y}_i}$ is the input data and $\mathbf{H_{c_i}}$ is the target data.
$\mathbf{\tilde{Y}_i} \in \mathbb{C}^{N_r \times N_t}$ are the received signal matrices collected through (10). $\mathbf{H_{c_i}} \in \mathbb{C}^{N_r \times N_t}$ are the corresponding true channel matrices collected through (4). $\mathbf{\tilde{Y}_i}$ is fed into CNN to approximate the corresponding true channels $\mathbf{H_{c_i}}$.\\
We use $N_t$ = 32 and $N_r$ = 16 as typical example for mmWave massive MIMO system.The CNN receives the tentatively estimated complex channel
matrix $\mathbf{\tilde{Y}_i} \in \mathbb{C}^{16 \times 32}$ as the input and separates their real and imaginary parts so that two 16 x 32 real-valued matrices are obtained. In the subsequent convolutional layer, the two matrices are processed by 64, 3 x 3 x 2 convolutional filters with the rectified linear unit
(ReLU) activation function to generate 64, 16 x 32 real-valued
matrices. Zero padding (ZP) is used when processing each
feature matrix so that its dimension remains unchanged
after convolution. For the
next four ZP convolutional layers, each uses 64, 3 x 3 x 64 convolutional filters and outputs 64, 16 x 32 real valued feature matrices. Then a batch normalization (BN) layer is added to avoid the gradient diffusion and overfitting. ReLU activation function is applied
for these four layers. The output layer uses two 3 x 3 x 64 convolutional filters to process the 64, 16 x 32 real-valued feature matrices and obtains the estimated real and imaginary parts of the channel matrices. After 
combining the corresponding real and imaginary parts, the
16 x 32 complex-valued estimated channel matrices, $\mathbf{\hat H_{c_i}}$ is obtained.\\ 
The objective of the training for the CNN is to minimize the MSE loss function
\begin{equation}
{MSE}_{loss} = \frac{1}{N_{tr}} \sum_{i = 1}^{N_{tr}}||\mathbf{H_{c_i} - \hat{H}_{c_i}}||^2_F
\end{equation}
where, $N_{tr}$ is the total number of samples in the training set and $||.||_F$ represents the Frobenius norm. \\

\section{Numerical Illustration and Results}
This section presents the simulation results to illustrate the performance of the proposed technique.
The number of antennas at the transmitter, $N_t$ = 32, the number of antennas at the receiver, $N_r$ = 16, number of cluster $N_c$ = 8 each cluster having two users, and the numbers of RF chains at the transmitter and the receiver $N_t^{RF} = N_r^{RF}$ = 2. $\mathbf F$ and $\mathbf W$ are set as the first $M_t$ columns of an $N_t \times N_t$ discrete Fourier transform (DFT) matrix and the first $M_r$ columns of an $N_r \times N_r$ DFT matrix. $M_r$ and $M_t$ are set as 2 and 32 respectively. The carrier frequency, $f_c$ = 28 GHz and the number of main paths is set as $L$ = 3. \\
For the CNN, the training set, validation set, and testing set contain 8100, 900, and 1900 samples respectively. Adam is used as the optimizer. The epochs are set as 100 and the corresponding learning rates is $3\times 10^{-4}$. The batch size is 128.\\
Normalized Mean Square Error (NMSE) is used to measure the performance of the channel estimation and is defined as:
\begin{equation}
\text{NMSE} = \mathbb E \Big\{\frac{\mathbf{||H_{c} - \hat{H}_{c}}||^2_F}{\mathbf{||H_{c}||^2_F}}\Big\}
%\mathbb \frac{{E{||\mathbf{H_{c_i} - \hat{H}_{c_i}}||^2_F}} {abc}}
\end{equation}
Fig. 3. shows the Normalized MSE performance versus signal-to-noise ratio (SNR) of CNN based channel estimation. The proposed technique is compared with Least Square estimate, MMSE estimate  and Cramer-Rao Bound (CRB). CRB represents the lower bound on the variance achievable by the estimator [14]. From the figure we can see that the proposed technique is close to the CRB and performs better than LS and MMSE estimator.
\begin{figure}[h]
	\centering
	\includegraphics[width=0.7\linewidth]{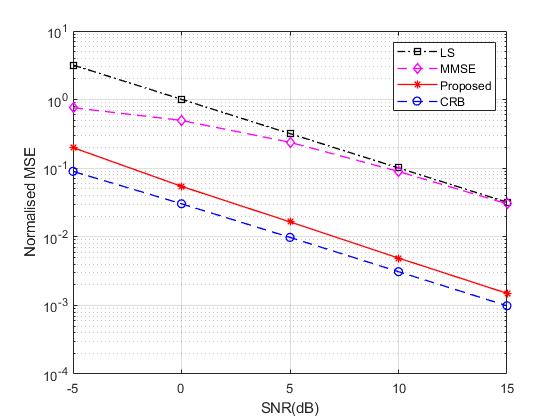} 
	\caption{Normalized MSE for CNN based Channel Estimation}
	%\label{fig:lena}
\end{figure}\\

Fig. 4. shows RMSE versus epochs for different learning rate for SNR = 15dB. From the figure it can be seen that RMSE decreases for learning rates $3 \times 10^{-2}$, $3 \times 10^{-3}$ and $3 \times 10^{-4}$. However when learning learning rate is further reduced to $3 \times 10^{-5}$, the model takes much longer time to converge for the same 20 epochs. Learning rate of $3 \times 10^{-4}$ gives better performance compared to other learning rates and hence chosen for our model.
\begin{figure}[h]
	\centering
	\includegraphics[width=0.7\linewidth]{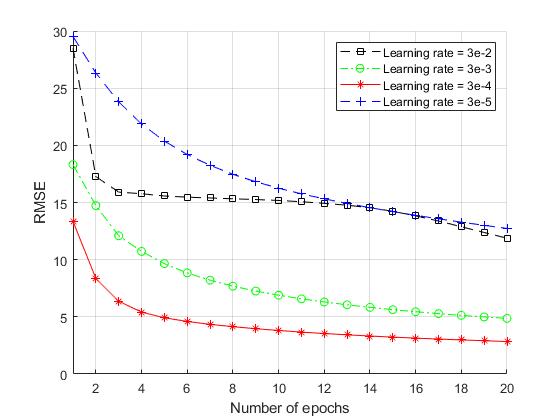} 
	\caption{RMSE vs Epochs for different learning rate for SNR = 15dB}
	%\label{fig:lena}
\end{figure}\\
Fig. 5. shows RMSE versus epochs for differnt number of convolutional layers (${L_c}$) for SNR=15dB. ${L_c}$ = 3 implies three covolutional layers (input layer, a hidden layer and output layer), ${L_c}$ = 4 implies four covolutional layers (input layer, two hidden layers and output layer) and so on. From the figure it can be seen that as the layer increases the model performance improves till ${L_c}$ = 6. When a hidden layer is added further i.e., ${L_c}$ = 7, the models takes longer time to converge for the same number of epochs. The model achieves best performance for ${L_c}$ = 6 and therefore selected for our model
\begin{figure}[h]
	\centering
	\includegraphics[width=0.7\linewidth]{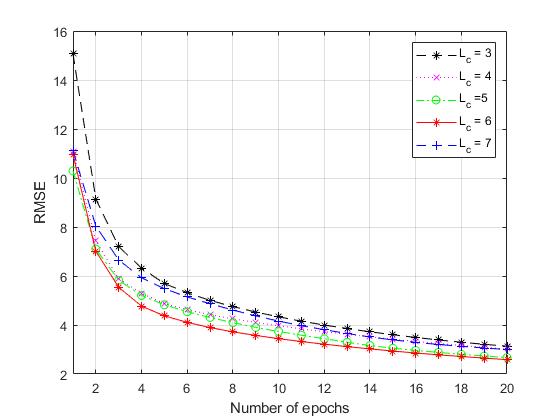} 
	\caption{RMSE vs for different number of convolutional layers for SNR = 15dB}
	%\label{fig:lena}
\end{figure}

%\newpage
\section{Conclusions}
A CNN based channel estimation approach for mmWave NOMA system is proposed. The proposed approach outperforms least square estimate, MMSE estimate and is close to the CRB. The proposed technique provides more accurate channel estimation.\\

\end{document}